# Time-Resolved Emission Study of a Thiophene-Modified Fluorescent Nucleoside in Solution and within Multiply-Modified Oligodeoxynucleotides

Mary Noe[1,☙], Yuval Erez[2,☙], Itay Presiado[3], Yitzhak Tor[1] and Dan Huppert[3]*

[1]*Department of Chemistry and Biochemistry, University of California, San Diego, La Jolla, California 92093-0358, United States*

[2]*Department of Physics of Complex Systems, The Weizmann Institute of Science, Rehovot 76100, Israel,*

[3]*Raymond and Beverly Sackler Faculty of Exact Sciences, School of Chemistry, Tel Aviv University, Tel Aviv 69978, Israel*

☙ These authors contributed equally to this work

*Corresponding author: Dan Huppert

e-mail: huppert@tulip.tau.ac.il

phone: 972-3-6407012

fax: 972-3-6407491

**Abstract**

Steady-state and time-resolved emission techniques were employed to study the photophysical properties of 5-(thien-2-yl)-2'-deoxyuridine (dU$^{Th}$), an isomorphic fluorescent nucleoside analog. We found that the emission lifetime of dU$^{Th}$ is dependent upon the solvent viscosity and obeys the Förster-Hoffman relation $I_f \propto \left(\frac{\eta}{T}\right)^{0.655}$ over a wide range of temperatures in 1-propanol, a glass-forming liquid. Upon incorporation into oligodeoxynucleotides, the average emission lifetime significantly increases, and the decay is non-exponential. We use a non-radiative decay model that takes into account the twist angle of the thiophene ring to fit the time-resolved emission decay curves.

**Introduction**

Fluorescence-based techniques have long served the biophysical community aiming to decipher the fundamental structural, folding and recognition features of biomolecules. Fluorescence-based tools have also been instrumental in advancing biophysical assays and high throughput screening techniques, particularly for drug discovery [1]. While numerous proteins contain fluorescent aromatic amino acids (e.g., tryptophan), or interact with fluorescent cofactors (e.g., NADH), nucleic acids present a challenge, as the native nucleobases are practically non-emissive [2,3]. Additionally, end labeling with common fluorescent tags (e.g., Cy dyes), although synthetically simple, is far from being optimal since: (a) the end labels are not necessarily sensitive to remote binding events, and (b) such common fluorophores are typically large, charged and frequently non-innocent [4]. These caveats have triggered the development of non-perturbing and responsive fluorescent nucleoside analogs, which have become powerful tools for investigating nucleic acids structure, dynamics and recognition [5,6].

Typically, a single native nucleoside is strategically replaced with its non-perturbing emissive counterpart [6]. Hybridization, folding, and recognition events are then monitored through changes in the photophysical characteristics of the emissive probe. A fundamentally distinct situation, which has only recently been investigated, is the replacement of multiple nucleosides within a single oligonucleotide with their emissive analogs [7,8]. Electronic and steric interactions between the emissive nucleoside as well as with the native nucleobases can significantly alter their photophysical features. Although the extent of electronic communication between probes embedded in the DNA p-stack is still debated, observations by Wagenknecht and Fiebig have suggested that small non-perturbing chromophores such as 2-aminopurine are only weakly electronically coupled [7]. Little time-resolved information is available, however, regarding new emissive nucleoside analogs, such as the 5-modified pyrimidines [8,9,10]. These visibly emitting and useful nucleosides can be viewed as molecular rotors, with free rotation about the carbon-carbon single bond linking the thiophene to the 5-position of the pyrimidine [10]. Therefore, they may be susceptible to neighbor–neighbor interactions, which has been recently explored through steady-state measurements [8]. To further shed light on such assemblies and their behavior, we report a thorough time-resolved photophysical analysis of an emissive nucleoside and the corresponding multiply-modified oligonucleotides.

Time-resolved studies of nucleosides and oligonucleotides (DNA and RNA) offer a window into their electronic properties, which may lead to a better understanding of damage and mutation. Eisinger and Shulman examined the photophysical properties of DNA, correlating the low fluorescence quantum yield of DNA with the short lifetime of the lowest excited state of a single nucleotide [11]. The fluorescence quantum yield of deoxynucleotides at room temperature was found to be in the order of $10^{-4}$, whereas the quantum yield increases to about 0.01 in liquid nitrogen. Recently femtosecond techniques were used to monitor the ultrafast non-radiative decay of nucleosides and DNA [12,13,14]. A review by Kohler and coworkers summarizes the recent achievements in the field [15]. In general, the fluorescence decay is bimodal, consisting of a major component of ultra-short decay-time (~300 fs) and a smaller amplitude component whose lifetime is longer (about 1 – 2 ps) for most of the nucleosides.

In the current study we employ both steady-state and time-resolved emission spectroscopy techniques to measure the fluorescence properties of a modified nucleoside, 5-(thien-2-yl)-2'-deoxyuridine ($dU^{Th}$) (Figure 1a) in a 1-propanol solution over a wide range of temperatures. Additionally, we explored the time-resolved behavior of $dU^{Th}$ upon incorporation into three distinct oligonucleotides in aqueous solution at room temperature. We found that the emission intensity and the emission decay time of $dU^{Th}$ increases as the 1-propanol solution temperature decreases. The non-radiative constant scales as $\left(\dfrac{\eta}{T}\right)^{0.655}$, where $\eta$ is the solution viscosity. We also found that in oligonucleotides, the emission of $dU^{Th}$ depends on the oligonucleotide sequence.

**Experimental Section**

Time-resolved fluorescence was acquired by using the time-correlated single-photon counting (TCSPC) technique, the method of choice when sensitivity, large dynamic range, and low-intensity illumination are important criteria in fluorescence decay measurements. For excitation, we used a cavity dumped Ti:Sapphire femtosecond laser, Mira, Coherent, which provides short, 80 fs, pulses of variable repetition rates, operating at the THG frequency, at the spectral range of 260 – 290 nm with a relatively low repetition rate of 800 kHz. A low rate may be important to excite dU$^{Th}$. The TCSPC detection system is based on a Hamamatsu 3809U photomultiplier and Edinburgh Instruments TCC 900 computer module for TCSPC. The overall instrumental response was about 35 ps (fwhm). The excitation pulse energy was reduced to about 10 pJ by neutral density filters. The large dynamic range of the TCSPC system (more than four orders of magnitude) enabled us to accurately determine the non-exponential photoluminescence decay profiles of dU$^{Th}$ fluorescence.

**Results**

### Steady-State Excitation and Emission of dU$^{Th}$

Figure 1 shows the excitation and emission spectra of dU$^{Th}$ at room temperature in pure water, dioxane and mixtures thereof. In previous studies, the emission intensity of this and related nucleosides was demonstrated to partially depend on the solvent viscosity [10]. The emission intensity of dU$^{Th}$ follows the Förster-Hoffman relation, $I_f \propto \left(\dfrac{\eta}{T}\right)^{0.655}$. The fluorescence quantum yield is in the range of one to two percent for solvent with viscosity around 1 cPoise. In the current study, we employ time-resolved emission techniques to further explore the fluorescence properties of dU$^{Th}$.

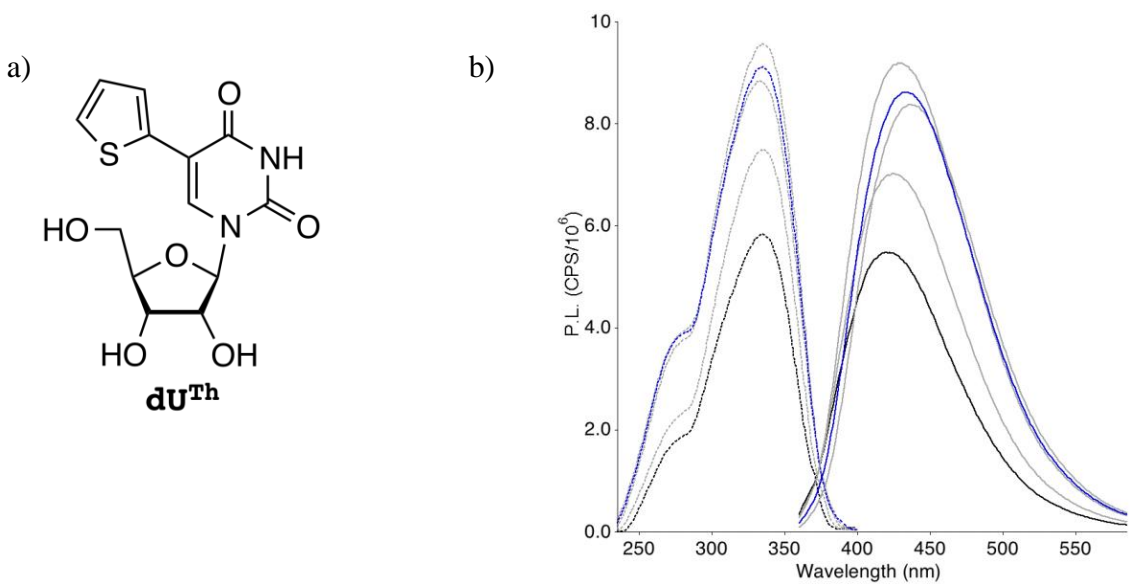

**Figure 1.** a) Structure of dU$^{Th}$; b) Excitation and emission spectra of dU$^{Th}$ in water (blue), dioxane (black) and mixtures of the two solvents (grey) at 294 K.

## Time-Resolved Emission of dU$^{Th}$

### Temperature Dependence of dU$^{Th}$ in 1-Propanol

Figure 2 shows on a linear scale, the time-resolved emission signals of dU$^{Th}$ in 1-propanol, measured at 222 K. The fluorescence was measured by the TCSPC technique with an instrument response function of ~35 ps. The sample was excited at 275 nm at a relatively slow repetition rate of 800 kHz. The time-resolved signals in 1-propanol are non-exponential at all of the emission wavelengths; the longer the wavelength, the smaller the deviation from exponential behavior. The average decay time increases with increasing wavelength.

We were able to fit most of the decay curves by a single stretched exponent, $\exp\left[-(t/\tau)^\alpha\right]$. The value of the stretch factor, α, falls as the wavelength decreases. For λ = 450 nm we found a good fit for τ = 1.5 ns and α ~ 0.77. Table 1 provides the stretched exponent fit of dU$^{Th}$ in 1-propanol at 222 K. At wavelengths of λ ≥ 510 nm, the signal fit also includes a small amplitude of a component characterized by a rise time.

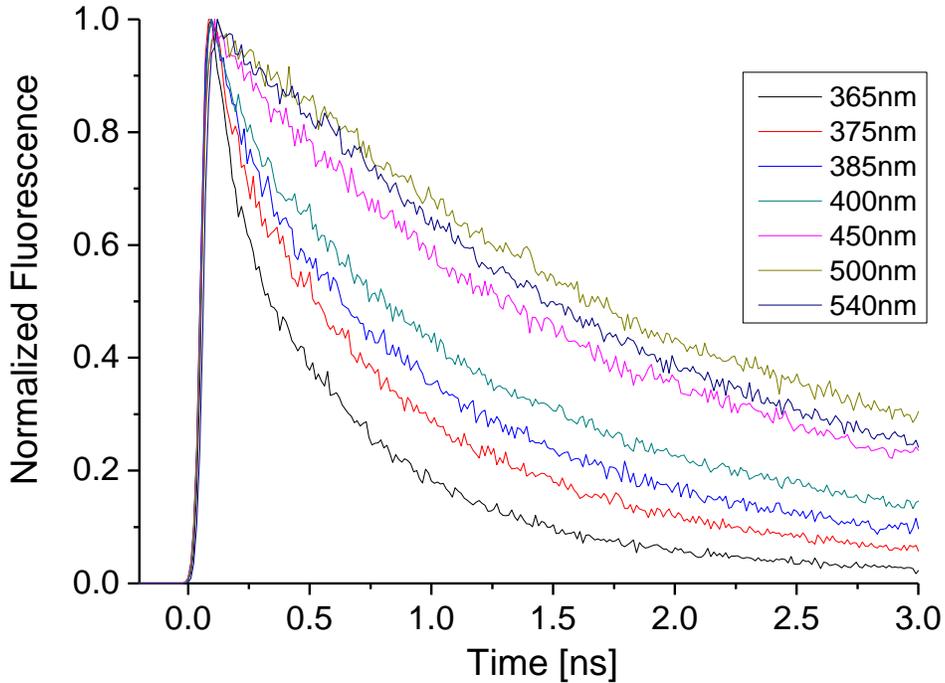

**Figure 2.** Wavelength dependence of the time-resolved emission of dU$^{Th}$ in 1-propanol, measured at 222 K.

1-Propanol is a glass-forming liquid, and at about 165 K the slope of the Arrhenius plot of $\log(\eta)$ vs 1/T increases as the temperature is lowered. The dielectric relaxation of 1-propanol shows temperature-dependence similar to that of the viscosity. Glass-forming liquids do not obey simple activation behavior as expressed by single activation energy, or as a constant slope of the dielectric-relaxation time, $\tau$, as a function of 1/T. Glass-forming liquids follow the empirical law of Vogel-Fulcher-Tammann (VFT):

$$\tau = \tau_0 \cdot \exp[DT_0/(T-T_0)], \tag{1}$$

or in its logarithmic form:

$$\log(\tau) = A + B/(T-T_0) \tag{2}$$

where A, B and $T_0$ are temperature-independent constants. The VFT fitting parameters of 1-propanol are: A = –10.53, B = 385.6 K and $T_0$ = 70.2 K [16]. The viscosity follows the dielectric–relaxation temperature dependence, and so the viscosity at 88 K should be greater by more than 12 orders of magnitude than that at room temperature ($\eta$ = 1.7 cP). For practical purposes, the high viscosity at 88K should prevent the internal rotation of the two ring systems with respect to each other, within the excited-state lifetime of about 7.5 ns. As seen in figure 2, the time-resolved emission decay of dU$^{Th}$ depends on the emission wavelength. The decay at short wavelengths is faster than at longer wavelengths. All the decay curves show a concave shape on a semilogarithmic scale, including the decay curve measured at 400 nm, which is located at the blue edge of the fluorescence spectrum.

Figure 3 shows, on a semilogarithmic scale, the time-resolved emission of dU$^{Th}$ measured at 400 nm at all of the measured temperatures over the range of 84 – 298 K. As seen in the figure, the rate of fluorescence decay decreases as the temperature is lowered. At temperatures T ≤ 160 K, the fluorescence decay is nearly temperature-independent. In the high-temperature region, solvent relaxation is relatively fast and this enables, within the excited-state lifetime, the relative ring rotation around the C-C' single bond. Rotation of the ring system around this bond brings the two ring systems into orthogonality. In this position the non-radiative rate reaches its maximum value. Thus, at high temperatures, the non-radiative rate constant $k_{nr}$ follows the temperature dependence of the solvent dielectric-relaxation rate, $1/\tau_D$. Table 2 provides the average fluorescence decay times of $dU^{Th}$ in 1-propanol at several temperatures in the range of 84 – 296 K.

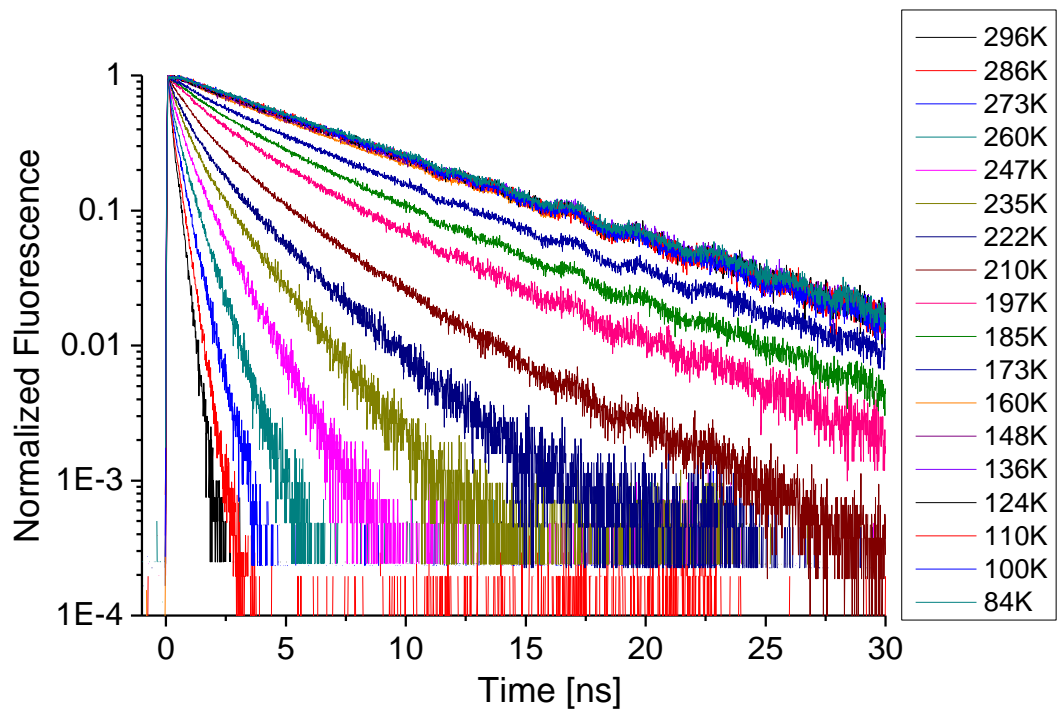

**Figure 3.** Time-resolved emission of $dU^{Th}$ measured at 400 nm in 1-propanol at several temperatures.

**Time-Resolved Emission of $dU^{Th}$ in Other Liquids**

Figure 4 shows the time-resolved emission of $dU^{Th}$ measured at 450 nm in acetonitrile, $H_2O$, methanol and ethanol. As seen in the figure, the average decay time increases with the solvent's viscosity. This phenomenon is characteristic of many floppy compounds, for which intramolecular rotation between two molecular moieties enhances the non-radiative decay rate. An example for such a non-radiative decay mechanism exists in thioflavin-T (ThT). Using ultrafast optical methods it was found that viscosity plays an important role in determining the non-radiative rate. The rate of the non-radiative decay process of ThT in 1-propanol varies by 4 orders of magnitude in the temperature range of 295 – 170 K. The signals measured at 450 nm show that the average decay time depends on the solvent viscosity. Table 6 gives the average lifetimes of the fluorescence decay curves shown in figure 4. As previously noted, the steady-state fluorescence quantum yield of many of such modified nucleosides follows the following relation [10]:

$$\Phi = B\left(\frac{\eta}{T}\right)^{\alpha} \tag{3}$$

For dU$^{Th}$, it was previously found that α = 0.65 ± 0.1.

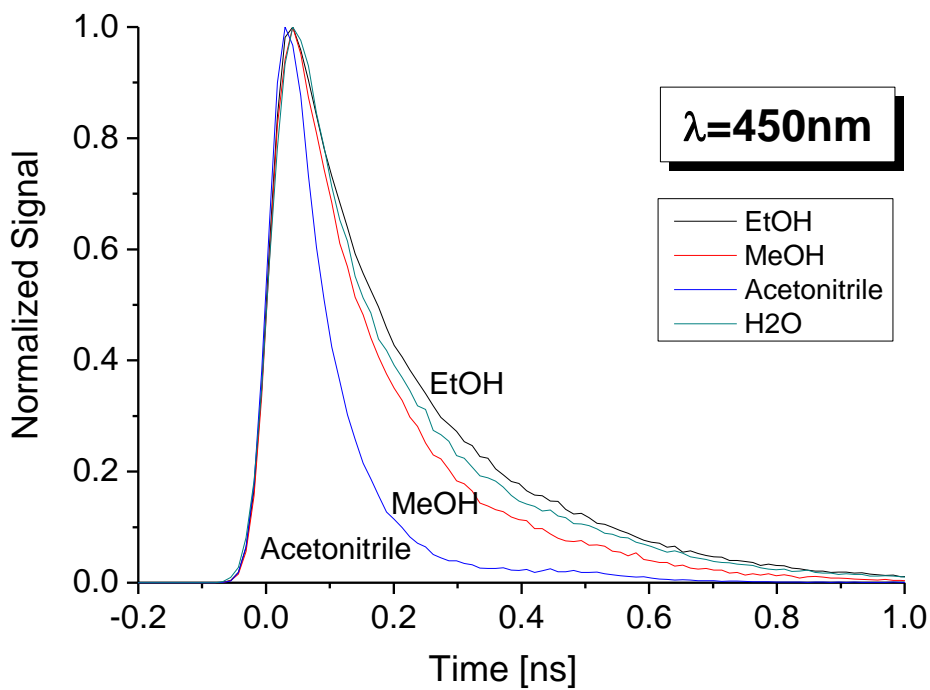

**Figure 4.** Time-resolved emission of dU$^{Th}$ in several liquids, measured at 450 nm.

**The Optical Properties of dU$^{Th}$ in Single-Stranded Oligonucleotides**

The steady-state fluorescence behavior of dU$^{Th}$ as an individual nucleoside differs significantly from those of oligonucleotides containing dU$^{Th}$. The molecular rotor behavior of dU$^{Th}$ suggests that it will display fluorescence sensitivity to its microenvironment upon incorporation into DNA, including the presence of adjacent fluorophores. This has been demonstrated through detailed steady-state studies [8]. To further explore these interactions and the impact of multichromophoric modifications, time-resolved fluorescence studies were performed on three previously synthesized oligonucleotides, each including either one or three fluorescent nucleoside analogs (See Figure 6) [8]. In designing these oligonucleotides, the highest possible sequence homology was maintained. Oligonucleotide **1** includes a single incorporation of dU$^{Th}$ near the center of the sequence. Oligonucleotide **2** includes three total incorporations of dU$^{Th}$ in central positions of the sequence, with each dU$^{Th}$ alternating with dA residues. Oligonucleotide **3** contains three adjacent incorporations of dU$^{Th}$.

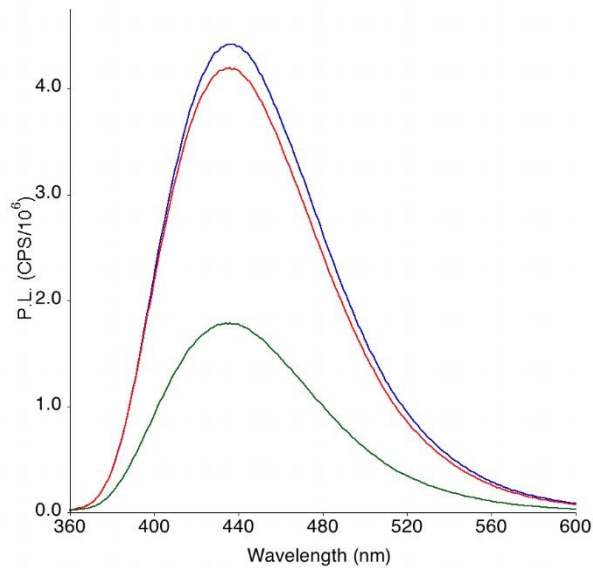

**Figure 5.** Steady state emission spectra of **1** (Blue), **2** (Red), and **3** (Green).

```
1  5' CCG GGA TAdUTh ATA GGC AG 3'
2  5' CCG GGA dUThAdUTh AdUThA GGC AG 3'
3  5' CCG GGA AdUThdUTh dUThAA GGC AG 3'
```

**Figure 6.** The 17mer oligonucleotide sequences containing the modified nucleoside. Sequences **1**–**3** contain either one or three incorporations of dU$^{Th}$.

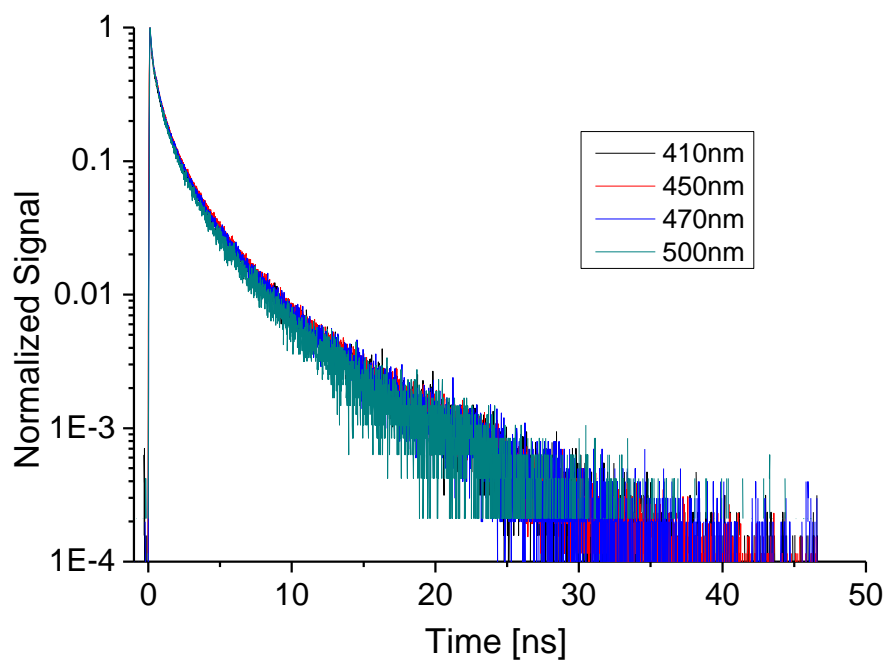

**Figure 7.** Time-resolved emission of dU$^{Th}$ in aqueous solution of oligonucleotide **2**, measured at several wavelengths.

Figure 5 shows the steady-state emission spectra of modified oligonucleotides **1**–**3**. Figure 6 shows the sequences of the three modified oligonucleotides. Note that although oligonucleotide **1** contains only a single incorporation of dU$^{Th}$, it has the highest fluorescence intensity. Interestingly, oligonucleotide **3**, with three adjacent incorporations of dU$^{Th}$, has the lowest fluorescence intensity. The drastic difference in fluorescence intensity between oligonucleotides **2** and **3**, both containing three modified nucleosides, is remarkable. This demonstrates a sequence-dependent fluorescence behavior, indicating that dU$^{Th}$ is sensitive to adjacent chromophores and fluorophores. To further probe these interactions, time-resolved measurements were explored.

Figure 7 shows the time-resolved emission of **2** at several wavelengths in the spectral range of 410 – 500 nm. The decay of **2** at all monitored wavelengths is nearly identical, and this is in contrast with the TCSPC signals of dU$^{Th}$ in water and other solvents. We attribute this observation to the fact that the thiophene ring angle in the oligonucleotide is fixed within the excited-state lifetime. The fluorescence decay profile is non-exponential with short, intermediate and long time decay components. Table 3 provides the fitting parameters of the three exponential fits Table 4 provide the fitting parameters of the stretched exponent fit. Interestingly, the stretch exponent fit is excellent, and it has only two free adjustable parameters, while the three exponent fit has 5 adjustable parameters. In the discussion section we provide a model explaining the non-exponentiality of the fluorescence decay and wavelength-independence of the decay profile of the molecule in the oligonucleotide.

Figures 8a and 8b show on linear and semilogarithmic scales the TCSPC time-resolved emission of the three oligonucleotides (**1–3**) as well as the free dU$^{Th}$ in water. As seen in the figure, the fluorescence decay strongly depends on the oligonucleotide into which dU$^{Th}$ is incorporated. The parameters of the multi-exponent fit and those of the stretched exponent fit of all three oligonucleotide samples are given in tables 3 and 4 respectively. The average emission lifetime of oligonucleotides **1**, **2**, and **3** are 2.27 ns, 0.95 ns and 1.72 ns respectively. The large differences in the decay times and profiles of dU$^{Th}$ are encouraging since they can be used as a sensor to gauge properties of single-stranded DNA. We further elaborate on this in the discussion section.

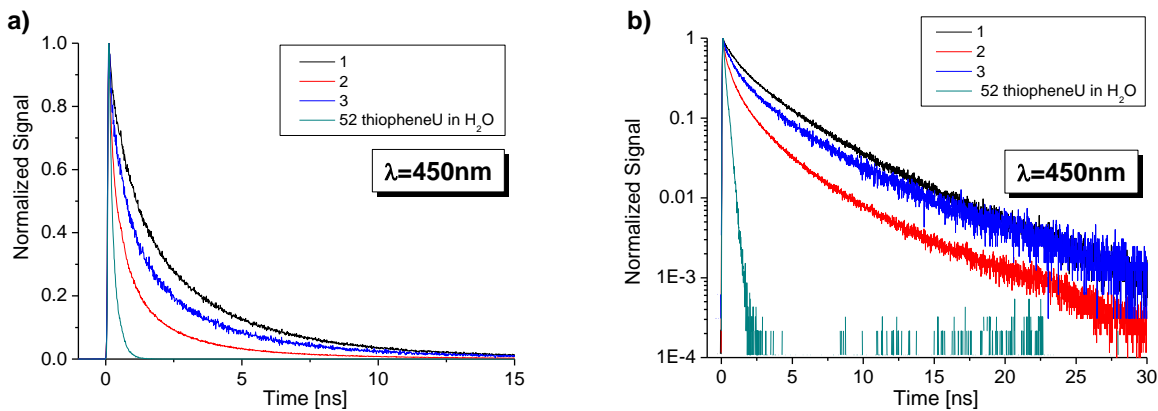

**Figure 8.** Time-resolved emission of **1**, **2**, **3**, and dU$^{Th}$ in water; a. linear scale; b. semilogarithmic scale.

**Discussion and Data Analysis**
**Temperature Dependence of the Non-Radiative Decay Process of dU$^{Th}$**

In this section we focus our attention on the qualitative relationship between the temperature-dependence of the dU$^{Th}$ non-radiative rate and the solvent properties, such as the viscosity and the dielectric-relaxation time. For this purpose we used 1-propanol, a protic solvent with a relatively high viscosity at room temperature with a marked temperature-dependence of the viscosity and dielectric relaxation. In our experiments, the viscosity is changed by lowering the temperature of the dU$^{Th}$ solution down to liquid-nitrogen temperatures. The dielectric relaxation of 1-propanol increases by 11 orders of magnitude as the temperature is reduced from room temperature to ~100 K. If the viscosity of 1-propanol with temperature follows the dielectric relaxation [17] then it also increases by about 11 orders of magnitude at ~100 K. In order to calculate the non-radiative rate constant, $k_{nr}$, from the experimental excited lifetime, we used the following procedure:

a. The non-radiative rate is calculated from the time-resolved emission measured at 400 nm, close to the steady-state emission-band maximum. At shorter wavelengths, the average decay time is shorter and the deviation from exponential decay is large. At longer wavelengths, λ > 500 nm, the time-resolved emission shows a small build-up time component, which increases the average lifetime.

b. We used the following relation to derive $k_{nr}$:

$k = k_r + k_{nr}$ (4)

where $k$ is the observed fluorescence-decay rate constant. For the pure radiative-decay rate constant, $k_r$, we used the average decay time of dU$^{Th}$ at the lowest temperature measured, $\tau_{av}$ = 7.4 ns at 88 K. We assume that at 88 K the non-radiative rate due to solvent relaxation and ring rotation around the C-C' bond is much smaller than 10 times $1/\tau_{av}$.

This approximation may not be correct and the pure radiative rate is probably even smaller, namely $1/k_r = \tau_r$ and $\tau_r$ = ~10 ns, as can be calculated using the Strickler-Berg equation [18] if the molar extinction coefficient and the shape of the absorption band are known. The existence of other non-radiative mechanisms, which are not related to the main one, also limits the effective radiative lifetime to $\tau'_r$ = ~7.4 ns.

Figure 9 shows the calculated non-radiative rate constant $k_{nr}$ of dU$^{Th}$ in 1-propanol as a function of 1/T. For comparison, the inverse of the dielectric-relaxation time, $1/\tau_D$, of 1-propanol is shown. We used the dielectric-relaxation measurements of Stickel and coworkers [17,19]. The dielectric relaxation of 1-propanol spans 12 orders of magnitude over the temperature range of 100 – 370 K. As previously mentioned, 1-propanol is a glass-forming liquid and does not crystallize but rather forms a glass at low temeopratures. This glass has liquid-like disorder, while molecular motions are strongly hindered. In a glass-forming liquid, the lower the temperature the larger the slope of the Arrhenius plot of the inverse of the dielectric–relaxation time, $1/\tau_D$, versus $1/T$. At roughly 165 K, the slope of $1/\tau_D$ of 1-propanol increases as the temperature decreases. The empirical VFT law, given by Equations (1) and (2), provides a good fit to the $1/\tau_D$ data of 1-propanol.

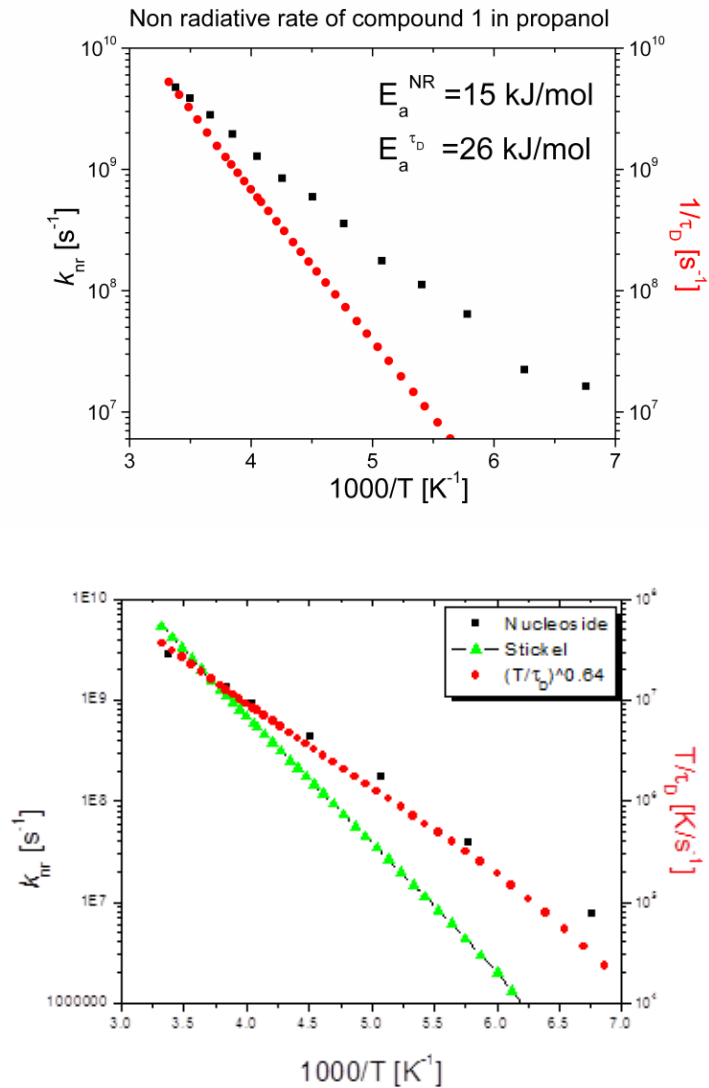

**Figure 9.** Non-radiative rate of dU$^{Th}$ in 1-propanol as a function of $1/T$. For comparison, $1/\tau_D$ of 1-propanol (see text).

As seen in Figure 9, there is a difference of the temperature dependence of $k_{nr}$ of dU$^{Th}$ in 1-propanol and that of $1/\tau_D$ of 1-propanol and $(T/\tau_D)^\alpha$, which was measured by Richert, Fischer and coworkers [17] over the temperature range of 125 – 295 K [17,19]. The value of $k_{nr}$ decreases by about two and a half orders of magnitude as the temperature drops to 165 K. The activation energies of both processes $k_{nr}$ and $1/\tau_D$ over this temperature range are about 15 and 25 kJ/mole respectively. We found that when $\alpha \simeq 0.655$, the slopes of $(T/\tau_D)^\alpha$ and of $k_{nr}$, both versus 1/T are about the same. At temperatures below 165 K, the non-radiative rate is smaller than the radiative rate by a factor of 10, which was measured at 88 K to be $1.35 \times 10^8$ s$^{-1}$ ($\tau_r \sim 7.4$ ns). As mentioned above, when $k_{nr} \ll k_r$ it is impossible to determine its value by steady-state or time-resolved spectroscopic techniques. It is interesting to note that the absolute values of $\tau_D$ and $\tau_{nr} = 1/k_{nr}$ show a rather small difference at room temperature. At a room temperature of 295 K, $\tau_D = 320$ ps and $\tau_{nr} = \sim 200$ ps.

**Inhomogeneous Kinetics Model.**

In recent papers [20,21], we used a model that accounts for inhomogeneous kinetics arising from a frozen structural medium surrounding an ensemble of excited molecules. With some modifications, the model is also applicable to kinetics in highly viscous solvents or semi-frozen matrices, such as the green fluorescent protein (GFP) chromophore in water-glycerol mixtures at low temperatures [22]. We wish to use the model to fit the time-resolved emission data of dU$^{Th}$ in oligonucleotides. The mathematical derivation of the inhomogeneous kinetics model for dU$^{Th}$ is similar to that of reference 20, which deals with the radiationless transition of the GFP chromophore in frozen water-glycerol solution. For simplicity, we assume that the distribution of the angle between the thiophene and thymine rings is Gaussian with a certain width defined by a variance $\sigma$ and a mean angle $\theta_0$. We shall use a continuous coordinate $\theta$ to define the distribution.

The distribution is given by

$$p(\theta) = \frac{1}{\sqrt{2\pi\sigma^2}} \exp\left[-\frac{(\theta-\theta_0)^2}{2\sigma^2}\right] \qquad (5)$$

where $\theta_0$ is the mean (the peak position) of the Gaussian. We assume that the rate coefficient of the non-radiative process depends exponentially on the coordinate $\theta$. We assume that the angle-dependence rate coefficient is given by

$$k_{nr}(\theta) = k_{nr}^0 \exp\left[-a|\theta - \theta_0|\right] \tag{6}$$

For convenience, the value of $\theta_0$ is zero. Since there is no difference between rotation to the right and rotation to the left, the substitution term in the exponent is taken in its absolute value.

In the static limit, where the dU$^{Th}$ conformation is time-independent with respect to the angle between the two ring moieties, the probability $P(t)$ that the excited molecule emits a photon by time $t$ after excitation is given by

$$P(t) = \exp(-t/\tau_f) \int_0^{\pi/2} p(\theta) \exp\left[-k_{nr}(\theta)t\right] d\theta \tag{7}$$

The first exponential accounts for the homogeneous radiative decay process, whereas the integral of the second exponential represents the non-radiative rate that depends on the angle distance distribution. The decay of $P(t)$ is non-exponential and depends on the excited-state radiative lifetime, $\tau_f$, $k_{nr}^0$, the parameter $a$ in equation 6, and the population Gaussian width $2\sigma^2$.

Figures 10a and 10b show the time-resolved emission signal of oligonucelotides **1**, **2**, and **3** on a linear and semi-logarithmic scales and the fit (solid line) using the inhomogeneous kinetics model. As seen in the figures, the fits are good for all three samples at short, intermediate, and long times. There are three adjustable parameters in the model. The fitting parameters of the three distinct oligonucleotides are given in table 5. The first parameter is the highest non-radiative rate coefficient, $k_{nr}^0$. Since we have no ground or excited potential curves, we cannot assign an angle between the two rings for $k_{nr}^0$. The value of $k_{nr}^0$ depends on the specific oligonucleotide. For **2**, **3**, and **1** we find $k_{nr}^0$ values of $1.5 \times 10^{10}, 5.4 \times 10^9$ and $3 \times 10^9$ s$^{-1}$ respectively. The second fitting parameter is the static population width, $2\sigma^2$, which was arbitrarily chosen to be 1 radian$^2$. We used the same value of $\sigma^2$ for all three oligonucleotides. The third adjustable parameter $a$ appears in the exponent in equation 6. The value of this parameter, 4.4 radian$^{-1}$, also remained invariant for all three oligonucleotides. This parameter means that the sensitivity of $k_{nr}$ to the angle is relatively high. A change of 1 radian in the angle decreases $k_{nr}$ by $e^{-4.4}$, i.e., nearly two orders of magnitude. There is an interplay between the population distribution width and the angle sensitivity exponential factor *a*. We can also get the same fit for a Gaussian Population distribution function P($\theta$) whose width is twice as narrow, and its value of *a* is twice as large. Therefore, there are only two adjustable parameters needed for a fit of the time-resolved emission signals are $k_{nr}^0$ and the standard deviation, $\sigma$. We intend to perform QM calculations of both the ground- and the excited-states potential energy curves as a function of the twist angle between the two rings. We expect these QM calculations to justify the inhomogeneous Gaussian distribution model.

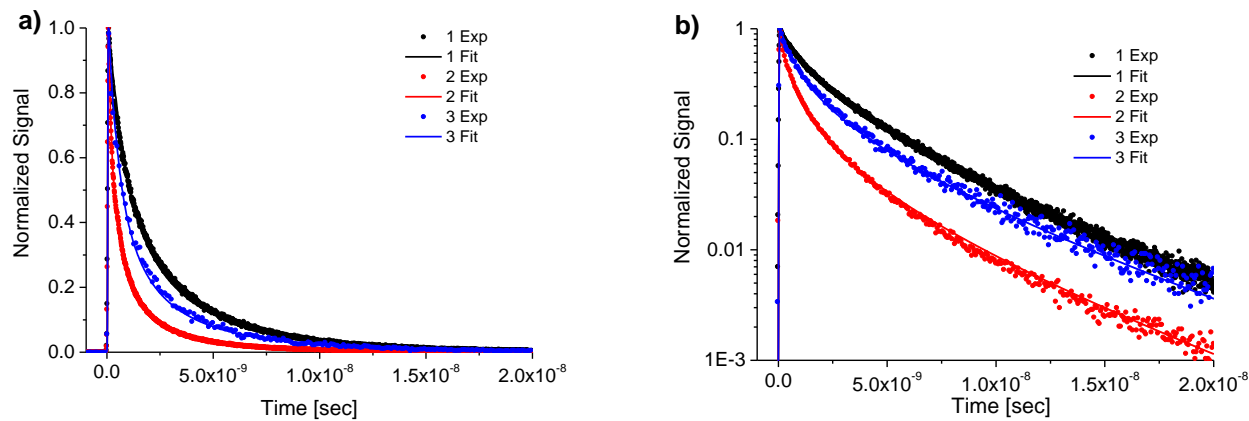

**Figure 10.** Fits to the inhomogeneous model of the time-resolved emission of three oligonucleotides (**1–3**). a. linearscale. b. semilogarithmic scale.

**Summary**

We used both steady-state and time-resolved emission techniques to study the photophysical properties of dU$^{Th}$, a modified fluorescent nucleoside. This nucleoside can be efficiently incorporated into oligonucleotides and thus can be used for investigating nucleic acid structure, dynamics and recognition [8]. Molecular rotors are molecules that are made of several moieties, where the rotation of one moiety with respect to the rest of the molecule forms a non-fluorescent twisted state. The fluorescence of molecular rotors is quenched by the rotation of the moiety, and therefore its intensity relies on the effective solvent viscosity in the micro-environment. We found that over a large range of temperatures, the fluorescence intensity and the average emission lifetime of dU$^{Th}$ in 1-propanol depends on solvent viscosity. The non-radiative rate coefficient follows the Hoffman-Förster relation with an exponent of 0.655 (i.e., $\left(\frac{\eta}{T}\right)^{0.655}$). We also studied the fluorescence properties of dU$^{Th}$ incorporated in three oligonucleotides, whose sequences are given in figure 6. We found a significant increase of the fluorescence intensity accompanied by extensive increase in average lifetime. Both the emission intensity and the lifetime depend on the specific sequence of the olgonucleotide. We used a model that explains the non-radiative decay process of dU$^{Th}$ incorporated into an oligonucleotide. The model assumes that the twist angle of dU$^{Th}$ with respect to pyrimidine ring controls the non-radiative decay rate. In an ensemble of dU$^{Th}$ molecules the twist angle is distributed normally. Using this model one can reasonably fit the time-resolved emission signals of dU$^{Th}$ in oligonucleotide.

**Acknowledgement**

We thank Ron Simkovitch for his help. This work was supported by grants from the James-Franck German-Israeli Program in Laser-Matter Interaction.

**Table 1. Stretched exponential fitting[a] of dU$^{Th}$ in 1-propanol ; T=222K**

| $\lambda$ [nm] | $\tau$ [ns] | $\alpha$ |
|---|---|---|
| 400 | 0.85 | 0.66 |
| 450 | 1.50 | 0.77 |
| 500 | 2.15 | 0.88 |

a. $I_f = \exp\left[-(t/\tau)^\alpha\right]$

**Table 2. Average fluorescence decay times of dU$^{Th}$ in 1-propanol as function of temperature**

| Temp [K] | $\tau_{av}$ [ns] | Temp [K] | $\tau_{av}$[a] [ns] |
|---|---|---|---|
| 296 | 0.21 | 185 | 4.14 |
| 286 | 0.25 | 173 | 5.18 |
| 273 | 0.34 | 160 | 6.61 |
| 260 | 0.48 | 148 | 6.88 |
| 247 | 0.71 | 136 | 7.18 |
| 235 | 1.03 | 124 | 7.08 |
| 222 | 1.38 | 110 | 7.14 |
| 210 | 2.05 | 100 | 7.11 |
| 197 | 3.27 | 84 | 7.35 |

a. $\tau_{av} = \int_0^\infty I_f(t)dt$

**Table 3a : Multi exponential fitting of time-correlated-single-photon-counting measurements of 1 at several wavelengths[a]**

| $\lambda$ [nm] | $a_1$ | $\tau_1$ [ns] | $a_2$ | $\tau_2$ [ns] | $a_3$ | $\tau_3$ [ns] | $\tau_{av}$[b] [ns] |
|---|---|---|---|---|---|---|---|
| 390 | 0.39 | 0.34 | 0.44 | 1.71 | 0.17 | 4.73 | 1.5 |
| 410 | 0.34 | 0.51 | 0.48 | 2.08 | 0.18 | 5.35 | 2.0 |
| 450 | 0.37 | 0.67 | 0.46 | 2.44 | 0.17 | 5.75 | 2.2 |
| 470 | 0.40 | 0.74 | 0.46 | 2.65 | 0.14 | 6.11 | 2.3 |
| 500 | 0.37 | 0.69 | 0.46 | 2.44 | 0.17 | 5.72 | 2.3 |

a. $y = \Sigma a_i \exp[-(t/\tau_i)]$.

b. $\tau_{av} = \int_0^\infty I_f(t)dt$

**Table 3b : Multi exponential fitting of time-correlated-single-photon-counting measurements of 2 at several wavelengths[a]**

| $\lambda$ [nm] | $a_1$ | $\tau_1$ [ns] | $a_2$ | $\tau_2$ [ns] | $a_3$ | $\tau_3$ [ns] | $\tau_{av}$[b] [ns] |
|---|---|---|---|---|---|---|---|
| 410 | 0.61 | 0.28 | 0.33 | 1.39 | 0.06 | 4.98 | 0.8 |
| 450 | 0.61 | 0.32 | 0.34 | 1.52 | 0.05 | 5.31 | 0.9 |
| 470 | 0.59 | 0.28 | 0.35 | 1.36 | 0.06 | 4.84 | 0.8 |
| 500 | 0.56 | 0.25 | 0.37 | 1.19 | 0.07 | 4.23 | 0.8 |

a. $y = \Sigma a_i \exp[-(t/\tau_i)]$.

b. $\tau_{av} = \int_0^\infty I_f(t)dt$

**Table 3c: Multi exponential fitting of time-correlated-single-photon-counting measurements of 3 at several wavelengths[a]**

| $\lambda$ [nm] | $a_1$ | $\tau_1$ [ns] | $a_2$ | $\tau_2$ [ns] | $a_3$ | $\tau_3$ [ns] | $\tau_{av}$[b] [ns] |
|---|---|---|---|---|---|---|---|
| 410 | 0.49 | 0.45 | 0.36 | 1.86 | 0.15 | 4.98 | 1.5 |
| 450 | 0.47 | 0.49 | 0.41 | 2.03 | 0.12 | 5.85 | 1.7 |
| 470 | 0.44 | 0.48 | 0.42 | 1.94 | 0.14 | 5.67 | 1.8 |
| 490 | 0.41 | 0.42 | 0.41 | 1.75 | 0.18 | 5.12 | 1.6 |

a. $y = \Sigma a_i \exp[-(t/\tau_i)]$.
b. $\tau_{av} = \int_0^\infty I_f(t)dt$

**Table 4. Oligomer data fitting parameters of using stretched exponent[a]**

|  | $dU^{Th,b}$ | 1 | 2 | 3 |
|---|---|---|---|---|
| $\tau$ [ns] | 0.45 | 0.36 | 1.86 | 0.15 |
| $\alpha$ | 0.49 | 0.41 | 2.03 | 0.12 |

a. $\exp\left[-(t/\tau)^\alpha\right]$
b. in 1-propanol at 222K

**Table 5. Oligomer data fitting parameters of inhomogenous model[a,b]**

|  | 1 | 2 | 3 |
|---|---|---|---|
| $k_r$ | $1.25 \times 10^8$ | $1.25 \times 10^8$ | $1.20 \times 10^8$ |
| $k_{nr}^0$ | $3.0 \times 10^9$ | $1.5 \times 10^{10}$ | $5.4 \times 10^9$ |

a. $a=4.44$
b. $\sigma^2=0.5$

**Table 6: The average lifetime of $dU^{Th}$**

|  | Ethanol | Methanol | Acetonitrile | $H_2O$ |
|---|---|---|---|---|
| $\tau$[ps] | 190 | 140 | 64 | 170 |

1. (a) A. J. Pope, U. M. Haupts, U. M., and K. J. Moore, Drug. Discov. Today **4**, 350, (1999). (b) R. P. Hertzberg, and A. J. Pope, Curr. Opin. Chem. Biol. **4**, 445 (2000). (c) M. Kimoto, R. S. Cox III, and I. Hirao, Exp. Rev. Mol. Diagn. **11**, 321 (2011).

2. Y. Tor, Tetrahedron **63**, 3425 (2007).

3. Accurate measurements reveal finite, albeit very small, fluorescence quantum yields for the natural nucleosides ($0.5 \times 10^{-4}$–$3 \times 10^{-4}$). This exceedingly weak emission is associated with very short excited state lifetimes (0.2–0.7 ps). See, for example, (a) M. Daniels, and W. Hauswirth, Science **171**, 675 (1971). (b) P. R. Callis, Annu. Rev. Phys. Chem. **34**, 329 (1983). (c) J, M. L. Pecourt, J. Peon, and B. Kohler, J. Am. Chem. Soc. **122**, 9348 (2000). (d) E. Nir, K. Kleinermanns, L. Grace, and M. S. de Vries, J. Phys. Chem. A **105**, 5106 (2001). (e) D. Onidas, D. Markovitsi, S. Marguet, A. Sharonov, and T. Gustavsson, J. Phys. Chem. B **106**, 11367 (2002).

4. (a) M. E. Sanborn, B. K. Connolly, K. Gurunathan, and M. Levitus, J. Phys. Chem. B **111**, 11064 (2007). (b) S. Ranjit, K. Gurunathan, and M. Levitus, J. Phys. Chem. B **113**, 7861 (2009). (c) B. J. Harvey, C. Perez, and M. Marcia, Photochem. Photobiol. Sci. **8**, 1105 (2009).

5. (a) C. Wojczewski, K. Stolze, and J. W. Engels, Synlett. **1999**, 1667 (1999). (b) M. E. Hawkins, Cell. Biochem. Biophys. **34**, 257 (2001). (c) M. J. Rist, and J. P. Marino, Curr. Org. Chem. **6**, 775 (2002). (d) D. P. Millar, Curr. Opin. Struct. Biol. **6,** 322 (1996). (e) C. J. Murphy, Adv. Photochem. **26**, 145 (2001). (f) A. Okamoto, Y. Saito, and I. Saito, J. Photochem. Photobiol. C: Photochem. Rev. **6**, 108 (2005). (g) R. T. Ranasinghe, and T. Brown, Chem. Commun. **2005**, 5487 (2005). (h) A. P. Silverman, and E. T. Kool, Chem. Rev. **106**, 3775 (2006). (i) J. N. Wilson, and E. T. Kool, Org. Biomol. Chem. **4**, 4265, (2006). (j) D. W. Dodd, and R. H. E. Hudson, Mini-Reviews in Organic Chemistry **6**, 378 (2009). (k) L. M. Q. Wilhelmsson, Rev. Biophys. **43**, 159 (2010).

6. R. W. Sinkeldam, N. J. Greco, and Y. Tor, Chem. Rev. **110**, 2579 (2010).

7. M. Rist, H.-A. Wagenknecht, and T. Fiebig, ChemPhysChem **3**, 704 (2002).

8. M. S. Noé, R. W. Sinkeldam, and Y. Tor, J. Org. Chem. **78**, 8123 (2013).

9. N. J. Greco, and Y. Tor, J. Am. Chem. Soc. **127**, 10784 (2005).

10. R. W. Sinkeldam, A. J. Wheat, H. Boyaci, and Y. Tor, ChemPhysChem **12**, 567 (2011).
11. J. Eisinger, and R. G. Shulman, Science **161**, 1311 (1968).
12. D. Onidas, D. Markovitsi, S. Marguet, A. Sharonov, and T. Gustavsson, J. Phys. Chem. B **106**, 11367 (2002).
13. D. Markovitsi, D. Onidas, T. Gustavsson, F. Talbot, and E. Lazarotto, J. Am. Chem. Soc. **127**, 17130 (2005).
14. D. Markovitsi, T. Gustavsson, and F. Talbot, Photochem.Photobiol. Sci. **7**, 717 (2007).
15. C. E. Crespo-Hernăndez, B. Cohen, M. P. Hare, and B. Kohler, Chem. Rev. **104**, 1977 (2004).
16. R. Richert, and C. A. Angell, J. Chem. Phys. **108**, 9016 (1998).
17. C. Hansen, F. Stickel, T. Berger, R. Richert, and E. W. Fischer, J. Chem. Phys. **107**, 1086 (1997).
18. S. J. Strickler, and R. A. Berg, J. Chem. Phys. **37**, 814 (1962).
19. R. Richert, F. Stickel, R. S. Fee, and M. Maroncelli, Chem. Phys. Lett. **229**, 302 (1994).
20. A. D. Kummer, C. Kompa, H. Niwa, T. Hirano, S. Kojima, and M. E. Michel-Beyerle, J. Phys. Chem. B **106**, 7554 (2002).
21. A. A. Voityuk, N. Rosch, M. Bixon, and J. Jortner, J. Phys. Chem. B **104**, 9740 (2000).
22. P. Leiderman, D. Huppert, and N. Agmon, Biophys. J. **90**, 1009 (2006).